\documentclass[11pt,a4paper]{article}
\usepackage{amsfonts,amssymb}
\usepackage[T2A]{fontenc}
\usepackage[cp1251]{inputenc}
\usepackage[english]{babel}
\usepackage{amsmath}%
\usepackage[unicode]{hyperref}

\newtheorem{theorem}{Theorem}
\newtheorem{lemma}{Lemma}

\def\const {\hbox{const}}

\numberwithin{equation}{section}

\title{\bf On a classification algorithm\\
 of the integrable two-dimensional lattices\\
  via Lie-Rinehart algebras}
\author{\bf I.T. Habibullin and   M.N. Kuznetsova}

\begin{document}
\maketitle



\abstract{In the article the problem of the integrable classification of nonlinear lattices depending on one discrete and two continuous variables is studied. By integrability we mean the presence of reductions of a chain to a system of hyperbolic equations of arbitrarily high order integrable in the Darboux sense. Darboux integrablity admits a remarkable algebraic interpretation: the Lie-Rinehart algebras related to both characteristic directions corresponding to the reduced system of the hyperbolic equations have to be of finite dimension. A classification algorithm based on the properties of the characteristic algebra is discussed. Some classification results are presented.}

\large

\section{Introduction}

We are interested in the problem of integrable classification of lattices of the form:
\begin{equation}  \label{eq0}  
u_{n,xy} = f(u_{n+1},u_n,u_{n-1}, u_{n,x},u_{n,y}), \quad -\infty < n < \infty,
\end{equation}
where the sought function $u_n = u_n(x,y)$ depends on the real $x,y$ and the integer $n$. The function  $f = f(x_1,x_2,\dots, x_5)$ of five variables is assumed to be analytic in a domain  $D\subset \mathbb{C}^5$.

The symmetry approach provides a very effective classification tool for integrable 1+1-dimensional models (see \cite{Adler}). However, for equations with three or more independent variables, higher symmetries contain non-local variables, and this circumstance causes serious technical problems that reduce the efficiency of the method \cite{Mikhailov98}. Various approaches to the investigation of integrable multidimensional models are discussed in the literature (see for instance, \cite{Bogdanov}--\cite{Sergeev}). For the purposes of an integrable classification, the method of reducing an equation to integrable two-dimensional models is often used. Usually, the authors require that the reduced models be soliton systems or integrable systems of the hydrodynamic type \cite{Fer-TMF} - \cite{Ferapontov2006}. In our opinion, the presence of a sufficient number of reductions that are Darboux integrable (see \cite{ShYam}-\cite{Yamilov}) can also be a sign of the integrability of a multidimensional model. In our recent study (see \cite{H2013}-\cite{HabPoptsovaUMJ}), we tested this idea by applying it to a nonlinear chain of the form (\ref{eq0}).

We have observed previously that any integrable lattice of the form (\ref{eq0}) admits the so-called degenerate cutting off boundary conditions. When such kind boundary conditions are imposed at two different points $n=N_1$ and $n=N_2$ then the lattice reduces to a Darboux integrable system of the hyperbolic type equations. We suggested and developed in our works \cite{H2013}-\cite{Kuznetsova19} a classification algorithm based on this observation. Let's briefly discuss the essence of the method.

We say that the constraint 
\begin{equation*} \label{cutoff}
u_0 = \varphi_0=\const
\end{equation*}
defines a degenerate boundary condition for the lattice \eqref{eq0} if it divides \eqref{eq0} into two independent semi-infinite lattices, defined on the intervals $-\infty<n<0$ and $0<n<+\infty$, respectively. There are two different kinds of the degenerate boundary conditions: regular and singular. In the regular case the point $(u_1, \varphi_0, u_{-1}, 0, 0)$ belongs to $D$ and the following conditions 
\begin{equation*}
\frac{\partial f(u_1, \varphi_0, u_{-1}, 0, 0)}{\partial u_1} = 0 \quad \mathrm{and} \quad \frac{\partial f(u_1, \varphi_0, u_{-1}, 0, 0)}{\partial u_{-1}} = 0
\end{equation*}
are met guaranteeing that for this choice of $u_0$ function $f$ does not depend on $u_1$ and $u_{-1}$, so that we have $f(u_1, \varphi_0, u_{-1}, 0, 0)=\hat{f}(\varphi_0)$. Therefore the equation \eqref{eq0} implies a relation
\begin{equation*}
\hat{f}(\varphi_0)=0,
\end{equation*}
from which the boundary value $\varphi_0$ is determined.

In the singular case we have as a rule two different constraints $u_0 = c_1$ and $u_0 = c_2$ with constants $c_1$, $c_2$ such that for the values $u_0 = c_1$ and $u_0 = c_2$ function $f(u_1, u_{0}, u_{-1}, u_{0,x}, u_{0,y})$ is not defined (the case $c_1 = c_2$ is not excluded), however the nearest equations
\begin{equation*}
u_{-1,xy} = f(c_1, u_{-1}, u_{-2}, u_{-1,x}, u_{-1,y})
\end{equation*}
and
\begin{equation*}
u_{1,xy} = f(u_2, u_1, c_2, u_{1,x}, u_{1,y})
\end{equation*}
are correctly defined, i.e. the functions $\tilde{f}(x,y,z,w) = f(c_1,x,y,z,w)$ and $\check{f}(x,y,z,w) = f(x,y,c_2,z,w)$ are analytic in some domains in $\mathbb{C}^4$.

\textbf{Example 1.} Evidently the Toda lattice equation
\begin{equation}
u_{n,xy} = e^{u_{n+1}-u_n} - e^{u_n - u_{n-1}}  \label{Toda}
\end{equation}
doesn't have any regular degenerate boundary condition. Its singular one is defined by two equations
\begin{equation}\label{-}
u_0 = -\infty
\end{equation}
and
\begin{equation}\label{+}
u_0 = +\infty.
\end{equation}
Indeed equation $u_{0,xy} = e^{u_1 - u_0}-e^{u_0 - u_{-1}}$ is not defined for $u_0 = \pm\infty$. Although close equations
\begin{equation*}
u_{-1,xy} = -e^{u_{-1}-u_{-2}}
\end{equation*}
and
\begin{equation*}
u_{1,xy} = e^{u_2-u_1}
\end{equation*}
obtained by setting $n=\pm1$ and $u_0=\pm\infty$ in (\ref{Toda}) are correctly defined. Thus the constraint (\ref{-}), (\ref{+}) divides the lattice \eqref{Toda} into two semi-infinite lattices
\begin{eqnarray*}
&&u_{-1,xy} = -e^{u_{-1}-u_{-2}}, \nonumber\\
&&u_{n,xy} = e^{u_{n+1}-u_n} - e^{u_n - u_{n-1}}, \quad -\infty<n<-1
\end{eqnarray*}
and
\begin{eqnarray*}
&&u_{1,xy} = e^{u_2 - u_1},\nonumber\\
&&u_{n,xy} = e^{u_{n+1}-u_n} - e^{u_n - u_{n-1}}, \quad 1<n<+\infty.
\end{eqnarray*}

\textbf{Example 2.} The lattice
\begin{equation}  \label{exp_latt}
u_{n,xy} = e^{u_{n+1} - 2 u_n + u_{n-1}}
\end{equation}
admits a singular degenerate cut-off $u_0 = -\infty$, since for this value of $u_0$ the r.h.s. of \eqref{exp_latt} implies that the closest equations take the form $u_{-1,xy} = 0$ and $u_{1,xy} = 0$.

\textbf{ Example 3.} Let us consider the lattice
\begin{equation*}
u_{n,xy} = u_{n,x}u_{n,y} \left(\frac{1}{u_{n+1}-u_{n}} - \frac{1}{u_n -u_{n-1}} \right)
\end{equation*}
found in \cite{Fer-TMF}, \cite{ShY}. It is easily proved that the lattice admits a regular degenerate cut off boundary condition
\begin{equation*}
u_0 =c
\end{equation*}
as well as the singular one
\begin{equation*}
u_0 = \infty.
\end{equation*}

\textbf{Example 4.} The lattices
\begin{equation*}
u_{n,xy} = (u_{n+1}-2u_n +u_{n-1})u_y\quad \mbox{and}\quad u_{n,xy} = (e^{u_{n+1}-u_n}-e^{u_n-u_{n-1}})u_y
\end{equation*}
admit regular degenerate boundary condition $u_0=c$, where $c$ is an arbitrary constant.

It is easy to verify that for any change of the variables $ u_n = g (v_n) $ applied to (\ref{eq0}), the degenerate boundary condition is again transformed into a degenerate one.

Inspired by these observations, we use the following definition of integrability in the article. 

{\bf Definition 1.} {\it The lattice \eqref{eq0} is called integrable if there exist boundary values $\varphi_0$ and $\varphi_1$ such that for any choice of the integer $N$, the hyperbolic type system 
\begin{eqnarray}
&&u_0 = \varphi_0, \nonumber \\
&&u_{n,xy}=f(u_{n+1},u_n,u_{n-1}, u_{n,x},u_{n,y}),\quad 1\leq  n \leq N, \label{eq00} \\
&&u_{N+1}=\varphi_{1} \nonumber 
\end{eqnarray} 
obtained from the lattice \eqref{eq0} is integrable in the sense of Darboux.}
 
We stress that the well-known integrable lattices of the form \eqref{eq0} considered in the Examples 1-4 are definitely integrable in the sense of the Definition 1 as well. 
 
Now we have to recall what Darboux integrability is. First we define the notion of the nontrivial $x$- and $y$-integrals. A function $I=I(x,\bar u,\bar u_x,\bar u_{xx},...)$ is called a $y$-integral if it satisfies the condition $D_yI=0$. Here $\bar u$ is a vector $\bar u=(u_1,u_2,\dots, u_N)$, $\bar u_x$ is its derivative and so on. Similarly, a function $J=J(y,\bar u,\bar u_y,\bar u_{yy},...)$ is an $x$-integral if it solves equation $D_xJ=0$.
Integrals of the form $I=I(x)$ and $J=J(y)$ are called trivial. A system \eqref{eq00} is called Darboux integrable if it admits a complete set of functionally independent integrals in both characteristic directions $x$ and $y$. Completeness means that the number of functionally independent integrals is $N$ in each direction.

Let $I=I(x,\bar u,\bar u_x,\bar u_{xx},...)$ be a nontrivial $y$-integral for the system (\ref{eq00}). We rewrite the equation $D_yI=0$ in the form 
\begin{equation}\label{Y}
YI=0
\end{equation}
where 
\begin{equation}\label{Ydef}
Y=\sum_{i=0} ^N \left(u_{i,y} \frac{\partial}{\partial u_i} + f_i \frac{\partial}{\partial u_{i,x}} + D_x(f_{i})\frac{\partial}{\partial u_{i,xx}} + \cdots  \right) \end{equation}
and $f_i=f(u_{i+1},u_i,u_{i-1},u_{i,x},u_{i,y})$.
Since the integral $I$ doesn't depend on the variable $u_{i,y}$ we have additional equations
\begin{equation}\label{Xj}
X_iI=0 \quad\mbox{where}\quad X_i=\frac{\partial}{\partial u_{i,y}}.
\end{equation}

In the sequel, we need the notion of characteristic algebra. Let $A$ denote the ring of locally analytic functions of the dynamical variables  $\bar u_{y},\bar u,\bar u_x,\bar u_{xx},\dots$. Consider the Lie algebra $L_y$ with the usual operation $[Z, W]=ZW-WZ$, generated by the differential operators $Y$ and $X_j$ defined in (\ref{Y}) and (\ref{Xj}) over the  ring $ A $, adding consistency conditions:
\begin{itemize}
\item[1).] $[Z,aW]=Z(a)W+a[Z,W]$,
\item[2).] $(aZ)b=aZ(b)$
\end{itemize} 
valid for any $Z,W\in L_y$ and $a,b\in A$. Roughly speaking, if $Z\in L_y$ and $a\in A$ then $aZ\in L_y$. In such a case algebra $L_y$ is called Lie-Rinehart algebra \cite{Rinehart}, \cite{Million}. We call it also characteristic algebra in the direction of $y$. In a similar way characteristic algebra $L_x$ in the direction of $x$ is defined.

Algebra $L_y$ is of a  finite dimension if it admits a  basis containing a finite number of the operators $Z_1,Z_2,\dots,Z_k\in L_y$ such that arbitrary element $Z\in L_y$ is represented as a linear combination of the form
$$Z=a_1Z_1+a_2Z_2+\dots a_kZ_k$$
where the coefficients are functions $a_1,a_2,\dots,a_k\in A$.

Due to the equations (\ref{Y}) and (\ref{Xj}) and according to the definition of the characteristic algebra an arbitrary $y$-integral  belongs to the kernel of any operator in $L_y$.
Moreover, the following statement is valid.

\begin{theorem} \label{Theorem 1}
System (\ref{eq00}) admits a complete set of the $y$-integrals (a complete set of the $x$-integrals) if and only if its characteristic algebra $L_y$ (respectively, characteristic algebra $L_x$) is of finite dimension.
\end{theorem}

\noindent
{\bf Corollary of Theorem 1.} {\it System (\ref{eq00}) is integrable in the sense of Darboux if both characteristic algebras $L_x$ and $L_y$ are of finite dimension.}

The remarkable work of A. B. Shabat \cite{Sh1995}, which gives a complete description of the characteristic Lie algebra for the Toda lattice \eqref{exp_latt}, is worth mentioning. We note that this was the first example of a characteristic Lie algebra for an equation with three independent variables.

\section{Some general properties of the characteristic algebras}

We apply the above algebraic integrability criterion to describe integrable cases of the lattice (\ref{eq0}). Since the elements of the characteristic algebras are vector fields with infinitely many components the problem of determining the dimensions of $L_y$, $L_x$  or of their subsets is a nontrivial task. To this end the lemma below can be used effectively \cite{ShYam, ZMHSbook}.

\begin{lemma} \label{lemma1}
If the vector field of the form
\begin{equation*} 
Z = \sum_{i=1}^N z_{1,i} \frac{\partial}{\partial u_{i,x}} + z_{2,i} \frac{\partial}{\partial	u_{i,xx}} + \cdots
\end{equation*}
solves the equation $\left[ D_x, Z \right] = 0$, then $Z=0$.
\end{lemma}

Let us evaluate the action of the operator $ad_{D_x}: Z\rightarrow[D_x,Z]$ on the basic operators in $L_y$.

\begin{lemma} \label{lemma2}
$[D_x,X_j]=-X_j(f_j)X_j,\quad  [D_x,Y]=-\sum _{j=1}^NY(f_j)X_j.$
\end{lemma}

Proof of Lemma 2. Evidently the operator of the total derivative $D_y$ acts on the set of the variables $\bar u_{y},\bar u,\bar u_x,\bar u_{xx},\dots$ according to the rule
\begin{equation}\label{Dyuy}
D_y=Y+\sum_{i=1}^{N}u_{i,yy}X_i,
\end{equation}
therefore we have $[D_x, D_y]=[D_x,Y+\sum_{i=1}^{N}u_{i,yy}X_i]=0$ or, the same
$$[D_x,Y]+\sum_{i=1}^{N}D_y(f_i)X_i+u_{i,yy}[D_x,X_i]=0.$$
After a transformation due to (\ref{Dyuy}) the latter implies
\begin{equation*}
[D_x,Y]+\sum_{i=1}^{N}Y(f_i)X_i+u_{i,yy}(X_i(f_i)+[D_x,X_i])=0.
\end{equation*}
By comparing the coefficients in front of the independent variables $u_{i,yy}$ we get the statement of the lemma.

\section{The first integrability condition}

Let us investigate the problem of describing lattices of the form (\ref{eq0}) integrable in the sense of Definition 1. 
We are looking for the function $ f $ using the system (\ref{eq00}) obtained by truncating the lattice  (\ref{eq0}). From the condition that the characteristic algebras $ L_x $ and $ L_y $ are finite-dimensional, differential equations are derived, which the function $ f $ must satisfy.

First we consider a sequence of the multiple commutators defined due to the rule
$$W_1=[X_0,Y],\quad W_{k+1}=[X_0,W_k] \quad \mbox{for} \quad k\geq1.$$
Specify the action of the operator $ad_{D_x}$ on the members of the sequence. For the first two of them we can easily find due to the Jacobi identity
\begin{eqnarray*}
&&[D_x,W_1]=-X_0(f_0)W_1-\sum_{j=1}^Np_{1,j}X_j,\\
&&[D_x,W_2]=-2X_0(f_0)W_2-X_0^2(f_0)W_1-\sum_{j=1}^Np_{2,j}X_j.
\end{eqnarray*}
It can be proved by induction that for the general value of $m$ these formulas look like
\begin{equation*}
[D_x,W_m]=-\sum_{k=0}^{m-1}C_m^kX_0^{m-k}(f_0)W_{k+1}-\sum_{j=1}^Np_{m,j}X_j
\end{equation*}
where the factors $C_m^k=\frac{m(m-1)\dots(m-k+1)}{k!}$ coincide with the binomial coefficients, factors $p_{m,j}$ are functions belonging to the ring $A$.

Since the characteristic  algebra $L_y$ is a finite dimensional linear space over the ring $A$ then there exists a natural $N$ such that $W_{N+1}$ is linearly expressed through the previous members of the sequence $W_1,W_2,\dots,W_N$ which are supposed to be linearly independent. It is easily checked that $W_1$ doesn't vanish, thus we get 
\begin{equation}\label{linearcom}
W_{N+1}=\lambda_NW_N+\lambda_{N-1}W_{N-1}+\dots+\lambda_1W_1, \quad N\geq1.
\end{equation}

Let us apply the operator $ad_{D_x}$ to both sides of (\ref{linearcom}) and obtain
\begin{equation*}
-(N+1)X_0(f_0)W_{N+1}-\frac{(N+1)N}{2}X_0^2(f_0)W_N=D_x(\lambda_N)W_N-\lambda_NNX_0(f_0)W_N+\dots,
\end{equation*}
where the tail contains only the combinations of the operators $W_{N-1}, W_{N-2},...W_1, X_1,...X_N$. Now we express $W_{N+1}$ due to the expansion (\ref{linearcom}) and then collect the coefficients in front of $W_N$:
\begin{equation}\label{DxlambdaN}
D_x(\lambda_N)+\lambda_{N}X_0(f_0)+\frac{(N+1)N}{2}X_0^2(f_0)=0.
\end{equation}
For simplicity we denote $\lambda:=\lambda_N$ and $\epsilon:=\frac{(N+1)N}{2}$ and omit the subindex $0$ in the expressions $u_0$ and $f_0$ when it does not lead to misunderstanding. Then (\ref{DxlambdaN}) takes the form
\begin{equation}\label{Dxlambda}
D_x(\lambda)+\lambda f_{u_y}+\epsilon f_{u_y,u_y}=0.
\end{equation}
Since the components of the vector fields $W_j$ depend on the variables $\bar u_{y},\bar u,\bar u_x,\bar u_{xx},\dots$ the factors $\lambda_j$ might depend only on these variables as well. Therefore,  (\ref{Dxlambda}) implies 
\begin{equation}\label{Dx}
\sum_{j=1}^Nf_j \frac{\partial\lambda}{\partial u_{j,y}}+u_{j,x}\frac{\partial\lambda}{\partial u_{j}}+u_{j,xx}\frac{\partial\lambda}{\partial u_{j,x}}+\dots+\lambda f_{u_y}+\epsilon f_{u_y,u_y}=0.
\end{equation}
By comparing the coefficients before $u_{j,xx}, u_{j,xxx}, ...$ in (\ref{Dx}) we show that the derivatives $\frac{\partial\lambda}{\partial u_{j,x}}$, $\frac{\partial\lambda}{\partial u_{j,xx}}, \dots$ all vanish and also we have $\frac{\partial\lambda}{\partial u_{j}}=0$ for $j\neq0$. Thus $\lambda$ depends only on $u$ and $u_y$ and it solves a system of the equations
\begin{equation}\label{system}
\frac{\partial\lambda}{\partial u_{x}}=0,\quad \frac{\partial\lambda}{\partial u}+\frac{f}{u_x} \frac{\partial\lambda}{\partial u_{y}}+ +\lambda\frac{f_{u_y}}{u_x}+\epsilon \frac{f_{u_y,u_y}}{u_x}=0.
\end{equation}

Let us reduce the system (\ref{system}) to the homogeneous form by introducing a new sought function $v(u_x,u,u_y,\lambda)$ such that $\lambda$ found from the equation $v(u_x,u,u_y,\lambda)=0$ solves system (\ref{system}). By differentiating equation $v(u_x,u,u_y,\lambda(u_x,u,u_y))=0$ with respect to the variables $u_x,u,u_y$ we obtain
$$\frac{\partial\lambda}{\partial u_{x}}=-\frac{\frac{\partial v}{\partial u_{x}}}{\frac{\partial v}{\partial \lambda}},\qquad 
\frac{\partial\lambda}{\partial u}=-\frac{\frac{\partial v}{\partial u}}{\frac{\partial v}{\partial \lambda}},\qquad 
\frac{\partial\lambda}{\partial u_{y}}=-\frac{\frac{\partial v}{\partial u_{y}}}{\frac{\partial v}{\partial \lambda}}. $$
After substitution of these formulas into (\ref{system}) we get 
\begin{eqnarray*}
&&l_1v:=\frac{\partial v}{\partial u_{x}}=0,\label{veq1}\\  \label{veq2}
&&l_2v:=\frac{\partial v}{\partial u}+\frac{f}{u_x}\frac{\partial v}{\partial u_{y}}-(\lambda\frac{f_{u_y}}{u_x}+\epsilon\frac{f_{u_yu_y}}{u_x})\frac{\partial v}{\partial \lambda}=0.
\end{eqnarray*}

Obviously, the desired function $ v $ is annihilated not only by the operators $ l_1 $ and $ l_2 $, but also by any operator from the Lie-Rinehart algebra $ L $ generated by these two operators. For instance, $v$ solves the equation
\begin{equation*}\label{veq3}
l_3v:=\frac{\partial}{\partial u_x}(\frac{f}{u_x})\frac{\partial v}{\partial u_{y}}-(\lambda\frac{\partial}{\partial u_x}(\frac{f_{u_y}}{u_x}) +\epsilon\frac{\partial}{\partial u_x}(\frac{f_{u_yu_y}}{u_x}))\frac{\partial v}{\partial \lambda}=0,
\end{equation*}
where $l_3=[l_1,l_2].$
We are interested in a solution $v$ which essentially depends on $\lambda$, i.e. it is supposed that $\frac{\partial v}{\partial \lambda}$ does not vanish identically, therefore dimension of the algebra $L$ must be no greater that three. This requirement can be called  {\bf the first integrability condition} in the direction of $y$. In a similar way we can derive an integrability condition in the direction of $x$. 

Thus we have two possibilities
\begin{itemize}
\item $\dim L=2, \quad l_3=0$;
\item $\dim L=3, \quad [l_1,l_2]=0,\quad [l_2,l_3]=0$.
\end{itemize}
The first case is realized only if $\frac{\partial}{\partial u_x}(\frac{f}{u_x})=0$. This equation gives immediately $$f(u_1,u,u_{-1},u_x,u_y)=g(u_1,u,u_{-1},u_y)u_x.$$ In the second case we obtain a linear equation for $\lambda$
\begin{equation*}
A \lambda + B =0,\label{AB}
\end{equation*}
where
\begin{eqnarray}
&&A = u_x (u_x f_{u_x}-f)f_{u u_x u_y} + f (u_x f_{u_x} - f) f_{u_x u_y u_y}+ u_x f_u f_{u_x u_y} + f f_{u_x u_y} f_{u_y} - \nonumber \\ \nonumber
&&- f_{u_y u_y} f^2_{u_x} u_x + f_{u_y u_y} f_{u_x} f - f f^2_{u_x u_y} u_x + f f_{u u_y} + f_{u_x} f_{u_y} f_{u_x u_y} u_x - f_{u_x} f^2_{u_y}-\\ \nonumber
&&- u_x f_{u_x} f_{u u_y} + f_{u_y} f_{u u_x} u_x - f_u f_{u_y} - f_{u_x u_y} f_{u u_x} u^2_x,
\end{eqnarray}
\begin{eqnarray}
&&B = u_x (u_x f_{u_x}-f) f_{u u_x u_y u_y} + f (u_x f_{u_x} - f) f_{u_x u_y u_y u_y} + 2 f f_{u_x u_y} f_{u_y u_y} -\nonumber\\ \nonumber
&&- f_{u_x} f_{u_x u_y} f_{u_y u_y} u_x + 2 u_x f_{u_y} f_{u_x} f_{u_x u_y u_y}  - f f_{u_y} f_{u_x u_y u_y} - u_x f^2_{u_x} f_{u_y u_y u_y} +\\\nonumber
&&+ f f_{u_x} f_{u_y u_y u_y}- u^2_x f_{u u_x} f_{u_x u_y u_y} + u_x f_u f_{u_x u_y u_y} - u_x f f_{u_x u_y} f_{u_x u_y u_y}\\ \nonumber
&&- f_{u_x} f_{u_y} f_{u_y u_y} + u_x f_{u u_x} f_{u_y u_y} - f_u f_{u_y u_y} - u_x f_{u_x} f_{u u_y u_y} + f f_{u u_y u_y}.
\end{eqnarray}
If $A$ doesn't vanish identically then we find $\lambda=-\frac{B}{A}$ and substitute it into the system  (\ref{system}) and also into the equation
\begin{equation*}\label{vveq3}
\frac{\partial}{\partial u_x}(\frac{f}{u_x})\frac{\partial \lambda}{\partial u_{y}}+\lambda\frac{\partial}{\partial u_x}(\frac{f_{u_y}}{u_x}) +\epsilon\frac{\partial}{\partial u_x}(\frac{f_{u_yu_y}}{u_x})=0
\end{equation*}
to derive equations that the function $ f $ must satisfy. If $A$ vanishes identically then $B$ vanishes as well and we get two equations 
\begin{equation}\label{vveq4}
A=0,\quad B=0.
\end{equation}

We have not investigated these rather complex equations. However we checked that the system (\ref{vveq4})  admits a solution of the form $f(u_1,u,u_{-1},u_x,u_y)=\alpha(u_1,u,u_{-1}) u_{n,x}u_{n,y} + \beta(u_1,u,u_{-1}) u_{n,x}+\gamma(u_1,u,u_{-1}) u_{n,y}+\delta(u_1,u,u_{-1})$.
In other words the lattice of the form
\begin{equation} \label{eq1} 
u_{n,xy}=\alpha u_{n,x}u_{n,y} + \beta u_{n,x}+\gamma u_{n,y}+\delta,
\end{equation}
satisfies the first integrability condition in both directions $x$ and $y$.

\section{Classification of a special case of the lattice \eqref{eq0}}

In this section we concentrate on the lattices of the form (\ref{eq1}) assuming that functions
\begin{equation} \label{eq11} 
 \frac{\partial\alpha(u_{n+1} ,u_n,u_{n-1} )}{\partial u_{n+1} }\quad \mbox{and} \quad\frac{\partial\alpha(u_{n+1} ,u_n,u_{n-1} )}{\partial u_{n-1} } 
\end{equation}
do not vanish identically\footnote{In the case when both of these requirements are violated  the lattice can be reduced to the form $$u_{n,xy}=\beta u_{n,x}+\gamma u_{n,y}+\delta.$$ On the classification of this kind lattices see \cite{Kuznetsova19}. }.

Operator $Y$ defined by the formula \eqref{Ydef} in this case may be represented as
\begin{equation}  \label{eq14} 
Y = \sum_{j=1}^N u_{j,y} Y_j + R,
\end{equation}
where 
\begin{eqnarray}
&&Y_i := \left[X_i, Y  \right] = \frac{\partial}{\partial u_i} + X_i(f_i)\frac{\partial}{\partial u_{i,x}} + X_i(D_xf_i)\frac{\partial}{\partial u_{i,xx}}+\cdots, \nonumber\\
&&R = \sum_{i=1}^N (\beta_i u_{i,x}+\delta_i)\frac{\partial}{\partial u_{i,x}}+ ((\alpha_i u_{i,x}+\gamma_i)(\beta_i u_{i,x}+\delta_i) + D_x(\beta_i u_{i,x}+\delta_i))\frac{\partial}{\partial u_{i,xx}}+\cdots \nonumber 
\end{eqnarray}
{\bf Remark 1.} {\it Since $\left[ X_j, Y_i \right] = 0$, $\left[ X_j, R \right] = 0$, we will restrict ourselves to the study of the subalgebra $\bar{L}_y$ of the characteristic algebra $L_y$ that is generated by the operators $R$ and $\left\{ Y_i \right\}_{i=1}^N$. It is obvious that in this case the algebras $ L_y $ and $ \bar {L} _y $ have a finite dimension only simultaneously.}

The main scheme we use below is to construct an appropriate sequence of the operators in the algebra and use the fact that a linear space spanned by the sequence over the ring $A$ must be of finite dimension. Since we are going to apply Lemma 1 the sequence $W_1, W_2, ...$ should be very special, it has to satisfy the condition
\begin{equation*}
\left[ D_x, W_k \right] = \sum_{j=1}^k r_j W_j, \quad r_j \in A.
\end{equation*}
Below in order to obtain a complete description of the integrable cases of the lattice \eqref{eq1}, \eqref{eq11} we use three sequences.

\subsection{First sequence}

We begin with the sequence defined as
\begin{equation*}  \label{eq15}
Y_0, \, Y_1, \, W_1 = \left[ Y_0, Y_1 \right], \, W_{k+1}=\left[ Y_0, W_k \right], \, k \geq 1.
\end{equation*}

We evaluated the action of the operator $ad_{D_x}$ on the members of the sequence:
\begin{eqnarray*}
&&\left[ D_x, Y_0 \right] = - a_0 Y_0, \quad \left[ D_x, Y_1 \right] = - a_1 Y_1, \\
&&\left[ D_x, W_1 \right] = -(a_0 + a_1)W_1 - Y_0(a_1)Y_1 + Y_1(a_0)Y_0,\\
&&\left[ D_x, W_k \right] = p_k W_k + q_k W_{k-1} + \cdots,
\end{eqnarray*}
where $a_i = \alpha_i u_{i,x} + \gamma_i$, $p_k = -(a_1 + k a_0)$,
\begin{equation*}
q_k = \frac{k-k^2}{2} Y_0(a_0) - Y_0(a_1)k.
\end{equation*}
Then we check that the operators $Y_0, Y_1, W_1$ are linearly independent and assume that for a  natural $M \geq 2$ operator $W_M$ is linearly expressed through the previous members of the sequence
\begin{equation}  \label{eq16}
W_M = \lambda W_{M-1} + \cdots
\end{equation}
while the operators $Y_0, Y_1, W_1, \ldots, W_{M-1}$ are linearly independent.
Now we apply the operator $ad_{D_x}$ to the equation \eqref{eq16} and get
\begin{equation*}
p_M W_M + q_M W_{M-1} + \cdots = D(\lambda) W_{M-1} + \lambda p_{M-1} W_{M-1}+ \cdots
\end{equation*}
Replace $W_M$ due to \eqref{eq16} and obtain the equation for determining $\lambda$
\begin{equation}  \label{eq17}
D_x(\lambda) = - a_0 \lambda - \frac{M(M-1)}{2} Y_0(a_0) - M Y_0(a_1).
\end{equation}

Since $\lambda = \lambda(u_0, u_1)$ we obtain from \eqref{eq17}
\begin{multline*}
\lambda_{u_0} u_{0,x} + \lambda_{u_1} u_{1,x} = -\left( \alpha_0 \lambda + \frac{M(M-1)}{2}(\alpha_{0,u_0}+\alpha^2_0) \right)u_{0,x} - \\
- M \alpha_{1,u_0} u_{1,x} - \left( \gamma_0 \lambda + \frac{M(M-1)}{2}(\gamma_{0,u_0} + \alpha_0 \gamma_0) + M \gamma_{1,u_0} \right).
\end{multline*}
Comparing the coefficients in front of the independent variables $u_{0,x}$ and $u_{1,x}$ we find a system of equations
\begin{eqnarray*} 
&&\lambda_{u_0}  = -\alpha_0 \lambda - \frac{M(M-1)}{2} (\alpha_{0, u_0}  + \alpha^2_0), \quad \lambda_{u_1}  = -M \alpha_{1,u_0},\label{eq18} \\
&&\gamma_0 \lambda + \frac{M(M-1)}{2}(\gamma_{0,u_0}+\alpha_0 \gamma_0 ) + M \gamma_{1,u_0} = 0. \label{eq19} 
\end{eqnarray*}
Solvability of the system provides an integrability condition for the lattice. It allows to specify the coefficient $\alpha$:
\begin{equation*} \label{eq20} 
\alpha(u_1,u_0,u_{-1}) = \frac{P'(u_0)}{P(u_0)+Q(u_{-1} )} + \frac{1} {M-1}  \frac{Q'(u_0)}{P(u_1)+Q(u_0)} - c_1(u_0).
\end{equation*}
Here $P$, $Q$ and $c_1$ are unknown functions of one variable, $M$ is unknown integer.

\subsection{Second sequence}

The next sequence is more complex, it contains three operators $ Y_0 $, $ Y_1 $, $ Y_2 $ and their multiple commutators:
\begin{gather}
Z_0 = Y_0, \, Z_1 = Y_1, \, Z_2 = Y_2, \, Z_3 = \left[Y_1, Y_0\right], \, Z_4 = \left[Y_2, Y_1\right], \nonumber\\ Z_5 = \left[Y_2, Z_3 \right], \, Z_6 = \left[ Y_1, Z_3\right], \, Z_7 = \left[Y_1, Z_4 \right], \, Z_8 = \left[Y_1, Z_5 \right].  \label{eq3_21} 
\end{gather}
Elements of the sequence $ Z_m $ for $ m> 8 $ are determined by the recurrence formula $ Z_ {m} = \left [Y_1, Z_ {m-3} \right] $. Note that this is the simplest test sequence generated by iterations of the map $ Z \rightarrow \left [Y_1, Z \right] $, which contains the operator $ \left [Y_2, \left [Y_1, Y_0 \right] \right] = Z_5 $.

\begin{lemma} \label{lemma3}
The operators $ Z_0, \, Z_1, \ldots Z_5 $ are linearly independent.
\end{lemma}

Actually the sequence splits down into three subsequences $ \{Z_ {3m} \} $, $ \{Z_ {3m + 1} \} $ and $ \{Z_ {3m + 2} \} $ and the latter is the most important.

\begin{theorem} \label{theorem2}
Assume that the operator $ Z_ {3k + 2} $ is represented as a linear combination
\begin{equation*} \label{eq3_25} 
Z_{3k+2}  = \lambda_k Z_{3k+1}  + \mu_k Z_{3k} + \nu_k Z_{3k-1}  + \cdots
\end{equation*}
of the previous members of the sequence \eqref {eq3_21} and none of the operators $ Z_ {3j + 2} $ for $ j <k $ is a linear combination of the operators $ Z_s $ with $ s <3j + 2 $. Then the coefficient $ \nu_k $ satisfies the equation
\begin{equation} \label{eq3_26} 
D_x(\nu_k) = -a_1 \nu_k - \frac{k(k-1)}{2}  Y_1 (a_1)- (k-1)Y_1 (a_0 + a_2).
\end{equation}
\end{theorem}
Proof of the theorem is based on the lemma.
\begin{lemma} \label{lemma4} 
Assume that all the conditions of Theorem~\ref {theorem2} are satisfied. Suppose that the operator $ Z_ {3k} $ (the operator $ Z_ {3k + 1} $) is linearly expressed in terms of the operators $ Z_i $, $ i <3k $ (respectively, $i < 3k+1$). Then in this expansion the coefficient at $ Z_ {3k-1} $ is zero.
\end{lemma}

Equation (\ref{eq3_26}) is equivalent to an overdetermined system of the equations for $\nu_k$
\begin{eqnarray*}
&&\nu_{k,u_0}  = -(k-1)\alpha_{0,u_1} , \label{eq3_32} \\
&&\nu_{k,u_1}  = -\alpha_1 \nu_k  - \frac{k(k-1)}{2}  (\alpha_{1,u_1}  + \alpha^2_1), \label{eq3_33} \\
&&\nu_{k,u_2}  = -(k-1)\alpha_{2,u_1}, \label{eq3_34} \\
&&\gamma_1 \nu_k  + \frac{k(k-1)}{2} \left( \gamma_{1,u_1} + \gamma_1 \alpha_1\right) + (k-1)(\gamma_{0,u_1} + \gamma_{2,u_1})=0.
\end{eqnarray*}
The compatibility condition of the system allows to prove that $ M = 2 $, $ k = 2 $. Both introduced above expansions are essentially specified 
\begin{equation} \label{eq4_2}
W_2 = \lambda W_1 + \sigma Y_1 + \delta Y_0,
\end{equation}
\begin{equation} \label{eq4_13}
Z_8 = \lambda Z_7 + \mu Z_6 + \nu Z_5 + \rho Z_4 + \kappa Z_3 + \sigma Z_2 + \delta Z_1 + \eta Z_0.
\end{equation}
The following  statement is valid
\begin{theorem}\label{theorem3}
The expansions (\ref {eq4_2}), (\ref {eq4_13}) take place if and only if the functions $ \alpha $, $ \gamma $ in the equation (\ref {eq1}) have the form:
\begin{eqnarray*}
&&\alpha(u_{n+1},u_n,u_{n-1}) = \frac{1}{u_n - u_{n-1}} - \frac{1}{u_{n+1}-u_n},\\
&&\gamma(u_{n+1},u_n,u_{n-1}) = r'(u_n) - r(u_n)\alpha(u_{n+1},u_n,u_{n-1}), \label{gamma_p3}
\end{eqnarray*}
where $r(u_n) = \frac{k_1}{2} u^2_n + k_2 u_n + k_3$ and the factors $k_i$ -- are arbitrary constants.
\end{theorem}

By similar reasonings in the $x$-direction we obtain an explicit expression for the coefficient $\beta$
\begin{equation*}
\beta(u_{n+1},u_n,u_{n-1}) = \tilde{r}'(u_n) - \tilde{r}(u_n)\alpha(u_{n+1},u_n,u_{n-1}),\label{eq3_61}
\end{equation*}
where $\tilde{r}(u_n) = \frac{\tilde{k}_1}{2} u^2_n + \tilde{k}_2 u_n + \tilde{k}_3$, and the coefficients $\tilde{k}_i$ -- are arbitrary constants.

\subsection{Third sequence}

The next step of our investigation is to refine the function $ \delta $. To do this, we build a new sequence on a set of multiple commutators containing the operator $R$ (see \eqref{eq14} above) that is a nonlocal part of the main characteristic operator $Y$:
\begin{equation*}
Y=\sum_{i=1}^N u_{i,y} Y_i + R.
\end{equation*}
Consider the following sequence of the operators in the characteristic algebra $\mathcal{L}(y,N)$:
\begin{gather*}
Y_{-1}, \, Y_0, \, Y_{1}, \, Y_{0,-1} = \left[Y_{0}, Y_{-1} \right], \, Y_{1,0} = \left[Y_1, Y_0 \right], \label{seq3}\\  
R_0=\left[Y_0, R \right], \, R_1 = \left[Y_0, R_0 \right], \, R_2 = \left[Y_0, R_1 \right], \, \ldots, \, R_{k+1}=\left[ Y_0, R_k \right]. \nonumber
\end{gather*}
By using the sequence we determine completely the desired coefficients of the quasilinear chain \eqref{eq1}
\begin{equation}  \label{res_latt}
u_{n,xy} =  \alpha_n u_{n,x} u_{n,y} + \beta_n u_{n,x} + \gamma_n u_{n,y} + \delta_n.
\end{equation}

\begin{theorem} \label{theorem4}
Up to point transformations there are three essentially different versions of the chain \eqref{res_latt} passing the necessary integrability test:

1) the chain \eqref{res_latt} reduces to the known 
Ferapontov-Shabat-Yamilov  chain (see \cite{Fer-TMF, ShY})
\begin{equation}  \label{var1}
u_{n,xy}=\alpha_nu_{n,x}u_{n,y},\quad \alpha_n = \frac{1}{u_n - u_{n-1}} - \frac{1}{u_{n+1}-u_n},
\end{equation}
degenerate boundary conditions are $u_0=c_0$, $u_{N+1}=c_1$ where $c_0, c_1$ are arbitrary constants;

2) 
\begin{equation}  \label{var2}
u_{n,xy}=\alpha_n(u_{n,x}u_{n,y}-u_n(u_{n,x}+u_{n,y})+u_n^2) +u_{n,x}+ u_{n,y} -u_n,
\end{equation}
degenerate boundary conditions are $u_0=0$, $u_{N+1}=0$;

3) 
\begin{equation}  \label{var3}
u_{n,xy}=\alpha_n(u_{n,x}u_{n,y}-s_{n}(u_{n,x}+u_{n,y})+s_{n}^2) + s'_{n}(u_{n,x}+u_{n,y}-s_n),
\end{equation}
where $ s_{n}=u_n^2+C$ and $C$ -- is an arbitrary constant.
Degenerate boundary conditions are $u_0=u_{N+1}=\sqrt{-C}$.
\end{theorem}
Equations 2) and 3) are found in our articles \cite{HabPoptsovaSIGMA17, HabPoptsovaUMJ}. It is proved in \cite{Poptsova19} that in the periodically closed case $u_n=u_{n+2}$ the lattice 2) admits higher symmetries.

\begin{theorem} \label{MainTheorem}
The lattices \eqref{var1}-\eqref{var3},  found in Theorem 4 are integrable in the sense of Definition 1 formulated in the Introduction.
\end{theorem}

Theorem can be proved by showing that the characteristic  algebra $L_y$ of the system of hyperbolic type equations obtained from lattice \eqref{res_latt} by imposing appropriate boundary conditions is of finite dimension (see \cite{HabPoptsovaUMJ}). To this end we show that the subalgebra $\bar L_y \subset L_y$ generated by the operators $\{Y_j \}_{j=1}^N$ and $R$ has a finite basis (see Remark 1 above)
$$R, \quad \{Y_i\}_{i=1}^{N},\quad \{Y_{i+1,i}\}_{i=1}^{N-1}, \quad \{Y_{i+2,i+1,i}\}_{i=1}^{N-2},\quad \ldots, Y_{N,N-1,\ldots,1}
$$
where $Y_{i+1,i}=[Y_{i+1},Y_i]$, $Y_{i+2,i+1,i}=[Y_{i+2},Y_{i+1,i}]$ and so on.

\section{Conclusions}

In the article a method for classification of integrable models with three independent variables is discussed. A conjecture is formulated that any integrable two dimensional lattice of the form \eqref{eq0} admits an infinite set of reductions being  Darboux integrable 1+1-dimensional systems of hyperbolic type equations (see Definition 1 above). The conjecture is approved by several examples. In \S3 for the lattice of general form \eqref{eq0} a necessary integrability condition is derived which might be useful for further investigations. In \S4 the efficiency of the algorithm is illustrated by presenting the results obtained earlier in \cite{HabPoptsovaSIGMA17}, \cite{HabPoptsovaUMJ}. We note that, in contrast to the symmetry classification, in the framework of this approach we do not use non-local variables.

\section{Conflict of Interest}

Conflict of Interest: ``The authors declare that they have no
conflicts of interest''.

Ismagil Talgatovich Habibullin (responsible for correspondence)\\
Institute of Mathematics, Ufa Federal Research Centre, Russian Academy of Sciences,\\ 112 Chernyshevsky Street,\\ Ufa 450008, Russian Federation \\ and
Bashkir State University,\\ 32 Zaki Validi Street, \\Ufa 450074, Russian Federation\\
Email {habibullinismagil@gmail.com}\\

Mariya Nikolaevna Kuznetsova\\
Institute of Mathematics, Ufa Federal Research Centre, Russian Academy of Sciences,\\ 112 Chernyshevsky Street, \\Ufa 450008, Russian Federation \\
Email {mariya.n.kuznetsova@gmail.com}\\

A list of keywords:

two-dimensional integrable lattice, $x$-integral, integrable reduction, cut-off condition, open chain, Darboux integrable system, characteristic Lie algebra.\\

\end{document}